\documentclass[11pt,a4paper]{article}
\usepackage[utf8]{inputenc}
\usepackage[english]{babel}
\usepackage{graphicx}
\usepackage{hyperref} 
\usepackage{geometry}
\usepackage{amsmath}
\usepackage{amssymb}
\usepackage{amsfonts}
\usepackage{multicol}
\usepackage{float}
\usepackage{caption}
\usepackage{subcaption}
\usepackage{cite}
\numberwithin{equation}{section}
\usepackage{amsthm}
\usepackage[utf8]{inputenc}
\usepackage{xcolor}
\newcommand{\gn}{G_{\rm N}}
\newcommand{\rs}{R_{\rm s}}
\newcommand{\rsa}{R_{\rm s}^{\ast}}
\newcommand{\lp}{l_{\rm p}}
\newcommand{\talp}{\tilde{\alpha}}
\newcommand{\tbet}{\tilde{\beta}}
\title{\bf Quantum corrected equations of motion in the interior and exterior Schwarzschild spacetime}
\author{Xavier Calmet$^{a}$\thanks{E-mail: X.Calmet@sussex.ac.uk},$\ $
Roberto Casadio$^{b,c}$\thanks{E-mail: Roberto.Casadio@bo.infn.it}$\ $ 
and Folkert~Kuipers$^{a,b}$\thanks{E-mail: F.Kuipers@sussex.ac.uk} 
\\
\\
{\em$^a$ Department of Physics and Astronomy, University of Sussex,}\\ 
{\em Brighton, BN1 9QH, United Kingdom}
\\
\\
{\em $^b$ Dipartimento di Fisica e Astronomia, Alma Mater Universit\`a di Bologna,}\\
{\em via Irnerio 46, 40126 Bologna, Italy}
\\
\\
{\em $^c$ I.N.F.N., Sezione di Bologna, I.S.~FLAG}\\
{\em viale B.~Pichat~6/2, 40127 Bologna, Italy}
}
\begin{document}
\maketitle
\vspace{5cm}
\begin{abstract}
In this paper we derive the leading quantum gravitational corrections to the geodesics and the equations
of motion for a scalar field in the spacetime containing a constant density star.
It is shown that these corrections can  be calculated in quantum gravity reliably and in a model independent way.
Furthermore, we find that quantum gravity gives rise to an additional redshift that results from the gradient
instead of the amplitude of the density profile. 
\end{abstract}
\thispagestyle{empty}
\pagebreak
\pagenumbering{arabic}
\section{Introduction}
In earlier work~\cite{Calmet:2019eof} we derived the leading quantum corrections to the interior and exterior
region of the spacetime containing a constant and uniform density star, which are classically described by the
well-known interior and vacuum Schwarzschild solutions.
These calculations were done in the framework of the effective field theory for quantum
gravity~\cite{Weinberg:1980gg, Barvinsky:1984jd, Barvinsky:1985an, Barvinsky:1987uw,
Barvinsky:1990up, Buchbinder:1992rb, Donoghue:1994dn}.
Corrections obtained in this way are the result of integrating out the quantum fluctuations of the graviton.

Remarkably, despite the fact that quantum general relativity is not renormalizable, it is possible to make
predictions in quantum gravity. These predictions apply to any model for which Lorentz invariance is a
fundamental symmetry, general relativity is the correct low energy limit, and for which quantum field theory
methods remain applicable up to the Planck scale.
The quantum gravitational effective action contains two parts consisting of local and nonlocal operators.
While the Wilson coefficients of the local part are non-calculable without knowing the ultra-violet complete
theory of quantum gravity, the Wilson coefficients of the nonlocal part of the action are calculable from
first principles and depend only on the infrared physics which is very well understood
as we know General Relativity.

Any unknown physics coming from an ultra-violet complete theory, would give rise to extra quantum corrections
in the form of local operators.
However, such physics only gives rise to contact interactions below the Planck scale.
For example, integrating out Kaluza-Klein interactions would give rise to contact interactions.
Furthermore, it was shown in~\cite{Calmet:2019eof} that corrections due to such contact interactions are
subleading in the case of a star, assuming higher order curvature terms are not unnaturally large.
The leading order corrections to the metric describing the spacetime around a star only depend
on the nonlocal physics which is calculable from first principles and in a model independent way,
without a detailed knowledge of the ultra-violet complete theory of quantum gravity.
\par
In this paper we will use the results from Ref.~\cite{Calmet:2019eof} to derive the leading quantum corrections
to the geodesics and the scalar waves in such a quantum corrected spacetime.
A complication in these calculations may arise, since the metric corrections and curvature invariants,
such as the Ricci scalar, diverge when the surface of the star is approached.
These secularities indicate a breakdown of the perturbative approach that is used, and result from the fact
that the interior Schwarzschild solution of general relativity contains a step-like discontinuity in the energy
density at the star surface.

Since the Einstein equations in general relativity only involve second order derivatives of the metric, 
step-like discontinuities result in acceptable $C^{1}$ metrics.~\footnote{Even Dirac delta-like discontinuities in the
energy density produce continuous metrics, which leads to the well-known case of shell-like sources.}
Quantum gravity in the effective field theory approach, on the other hand, is an infinite derivative theory
and it therefore requires $C^{\infty}$ sources in order to produce continuous metrics.
In other words, one should also determine a quantum correction to the matter source which makes it 
compatible with the effective quantum equations for the metric.
However, quantum corrections to the uniform matter source appear really necessary only within 
a layer of thickness of the order of the Planck length around the surface, and are expected to remain
phenomenologically negligible.

In any case, and although the non-smooth solutions of general relativity are not expected to be physical,
they can still serve as important toy models. 
This is particularly true for the Schwarzschild interior, as it is an analytical solution of the Einstein
equations, that could approximate compact objects.

While it seems difficult to find practical applications for our results, they are a further demonstration
that model independent calculations are possible in quantum gravity at energies below the Planck scale.
This is in sharp contrast to the standard lore which states that quantum gravity is a mystery:
we do not have a theory of quantum gravity and thus quantum gravitational calculations are not possible.
This is simply not true and our results help to reinforce this point.
As such, our findings are very important as they further demonstrate that quantum gravitational calculations
are possible at energies below the Planck scale.

This paper is organized as follows:
in the next section we state the results derived in~\cite{Calmet:2019eof};
in section~\ref{S:3}, we solve the radial geodesics perturbatively and derive the leading quantum corrections;
in section~\ref{S:4} we turn to the radial modes of the scalar field and solve their equations of motion perturbatively
to derive the leading quantum corrections;
finally in section~\ref{S:5} we conclude.
\section{The quantum corrected metric}
\label{S:2}
We here consider the quantum corrected metric derived in~\cite{Calmet:2019eof}, which
is static and spherically symmetric and can therefore be written as
\begin{equation}
\label{eq:metricgen}
	ds^2 = -f(r) \,dt^2 
	+ g(r)\, dr^2
	+ r^2\, d\Omega^2.
\end{equation}
where $d\Omega^2 = d\theta^2 + (\sin\theta)^2\, d\phi^2$.
Outside the star of radius $R_{\rm s}$ (that is, for $r>R_{\rm s}$), the metric functions are given by
\begin{align}
	f(r) &= 1 - \frac{2\,\gn M}{r} + \alpha_{\rm e}(r)
	\ ,
	\\
	g(r) &= \left(1 - \frac{2\,\gn M}{r} \right)^{-1} + \beta_{\rm e}(r)
	\ ,
\end{align}
where
\begin{align}\label{eq:QcorOut}
	\alpha_{\rm e}(r) &= \talp\, \frac{2\, \gn\, \lp^2\, M}{\rs^3}
	\left[ 2\, \frac{\rs}{r} + \ln \left( \frac{r-\rs}{r+\rs} \right) \right] 
	+\mathcal{O}\left(\gn^3\right)
	\ ,
	\nonumber
	\\
	\beta_{\rm e}(r) &= \tbet\, \frac{2 \,\gn\, \lp^2\, M}{r \,(r^2 - \rs^2)}
	+ \mathcal{O}\left(\gn^3\right)
	\ ,
\end{align}
with~\footnote{The values for $\alpha$, $\beta$ and $\gamma$ can be found in~\cite{Calmet:2019eof}.}
\begin{align}
	\talp &= 96\,\pi\, (\alpha + \beta + 3\gamma)
	\\
	\tbet &= 192\,\pi\,(\gamma - \alpha)
	\ .
\end{align}
In the stellar interior (given by $0\le r<R_{\rm s}$), we likewise have
\begin{align}
	f(r) &= \frac{1}{4}\left( 3\ \sqrt{1 - \frac{2\, \gn M}{\rs}} 
	- \sqrt{1 - \frac{2 \,\gn M r^2}{\rs^3}} \right)^2
	+\alpha_{\rm i}(r)
	\ ,
	\\
	g(r) &= \left(1 - \frac{2\, \gn M r^2}{\rs^3}\right)^{-1} + \beta_{\rm i}(r)
	\ ,
\end{align}
where now
\begin{align}\label{eq:QcorIn}
	\alpha_{\rm i}(r) &= \talp\, \frac{2\, \gn\, \lp^2\, M}{\rs^3}\,
	\ln \left(\frac{\rs^2}{\rs^2 -r^2} \right) + \mathcal{O}(\gn^3),\nonumber\\
	\beta_{\rm i}(r) &= \tbet\, \frac{2\, \gn\,\lp^2\, M\, r^2}{\rs^3\, (\rs^2 - r^2)}
	+ \mathcal{O}(\gn^3)
	\ .
\end{align}
Moreover, we assume throughout the paper that the Buchdahl limit~\cite{Buchdahl:1959zz} is satisfied,
so that
\begin{equation}
	\rs \geq \frac{9}{8}\, (2\,\gn\,M)
	\ .
\end{equation}
\par
Let us remark that the Newton constant $\gn$ is dimensionful and the displayed perturbation expansion
is therefore a shorthand notation for two contributions, which are different in nature.
In particular, 
\begin{equation}
	\mathcal{O}\left(\gn^3\right) 
	=
	\lp^2\, \mathcal{R} \, \mathcal {O}\left([2\gn M/\rs]^2\right) + \mathcal {O}\left(\lp^4\, \mathcal{R}^2\right)
	\ ,
	\label{OGn}
\end{equation}
where $\lp$ is the Planck length, and $\mathcal{R}$ is the curvature scalar.
The true perturbation parameters are thus the inverse of the radius of curvature in units of the Planck length
and the compactness of the star, which are dimensionless as they should.
\par
Furthermore, the quantum corrections become secular when $r\sim\rs$.
This secularity can be avoided, if the layer 
\begin{equation}
\label{eq:ExclInt}
	(1-\delta)\,\rs<r<(1+\delta)\,\rs
	\qquad {\rm with} \quad 
	\delta \sim \left(\frac{2\gn M}{\rs} \right) \left( \frac{\lp}{\rs}\right)^2
\end{equation}
is excluded, as discussed in~\cite{Calmet:2019eof}.
\par
Finally, we recall that the metric can be rewritten as
\begin{equation}
\label{eq:metricgen}
	ds^2 = f(r) (-dt^2 + dr_\ast^2)
	+ r^2 d\Omega^2
	\end{equation}
by introducing the tortoise coordinate
\begin{equation}
	r_\ast = \int^r \sqrt{\frac{g(r')}{f(r')}}\, dr'
	\ .
\label{eq:rast}
\end{equation}
This form is particularly useful fo studying waves and will be employed in section~\ref{S:4}.
\section{Geodesics}
\label{S:3}
Geodesic equations can be derived in a way similar to the derivation in a Schwarzschild metric.
The quantum corrected star metric has four Killing vectors.
Three of those are due to the spherical symmetry, and one due to time-invariance.
We use two of these Killing vectors to fix the direction of the angular momentum along the
polar axis by setting
\begin{equation}
	\theta=\frac{\pi}{2}
	\ .
\end{equation}
The remaining two Killing vectors can then be written as
\begin{align}
	K^\mu &= (\partial_t)^\mu
	\ ,
	\\
	R^\mu &= (\partial_\phi)^\mu
	\ ,
\end{align}
and can be used to define a conserved energy 
\begin{equation}
E = - K_\mu \, \frac{dx^\mu}{d\lambda} = f(r)\, \frac{dt}{d\lambda}
\end{equation}
and a conserved angular momentum
\begin{equation}
L = R_\mu\, \frac{dx^\mu}{d\lambda} = r^2\, \frac{d\phi}{d\lambda}
\ .
\end{equation}
Furthermore, along geodesics the quantity
\begin{equation}
\label{eq:geodesiceq}
	\epsilon = - g_{\mu\nu}\, \frac{dx^\mu}{d\lambda}\, \frac{dx^\nu}{d\lambda}
\end{equation}
is also conserved.
For massive particles we can set $\epsilon=1$, as long as we identify $\lambda=\tau$
as the proper time along the geodesic.
For massless particles $\epsilon=0$ with $\lambda$ an arbitrary affine parameter.
By making use of the conserved quantities, we can rewrite Eq.~\eqref{eq:geodesiceq} as
\begin{equation}
	\left( \frac{dr}{d\lambda}\right)^2 + \frac{1}{g(r)} \left(\frac{L^2}{r^2} + \epsilon\right) = \frac{E^2}{f(r)\,g(r)}
	\ .
\end{equation}
Compatibly with the quantum corrections described in section~\ref{S:2}, 
we will solve this equation perturbatively in the Planck length and the star compactness,
by writing
\begin{equation}
	r(\lambda) = r_{\rm c}(\lambda) + r_{\rm q}(\lambda)
	\ ,
\end{equation}
where
\begin{equation}
r_{\rm c}(\lambda) 
= \sum_{m=0}^{\infty}  r_{0,m}(\lambda)
	\left(\frac{2\,\gn M}{\rs}\right)^{m}
\end{equation}
represents the classical trajectory, and
\begin{equation}
	r_{\rm q}(\lambda) 
	= \sum_{n=1}^{\infty} \sum_{m=0}^{\infty}  r_{n,m}(\lambda)
	\left(\frac{\lp}{\rs}\right)^{2n}
	\left(\frac{2\,\gn M}{\rs}\right)^{m}
\end{equation}
is the quantum correction.
\subsection{Exterior region}
In the exterior region, $r>\rs$, we can write
\begin{equation}
	\left( \frac{dr}{d\lambda}\right)^2 
	+ \frac{L^2}{r^2} 
	- \frac{2\,\gn M}{\rs} \left(\frac{L^2}{r^2} + \epsilon \right) \frac{\rs}{r} 
	+ V_{\rm q}(r) 
	= \eta^2
	\ ,
\end{equation}
where $\eta = \sqrt{E^2 - \epsilon}$ and the effective quantum potential is given by
\begin{equation}
	V_{\rm q}(r) = E^2\, \alpha_{\rm e}(r) + \eta^2\, \beta_{\rm e}(r) - \frac{L^2}{r^2}\, \beta_{\rm e}(r)
	\ .
\end{equation}

We notice that the term proportional to $E^2$ signals a violation of the equivalence principle, 
since the acceleration undergone by the particle following the geodesic depends on its energy.
However, $\alpha_{\rm e}=\mathcal{O}(\gn^2)\sim (\lp/\rs)^2$ in the sense explained after Eq.~\eqref{OGn}, 
and the size of this violation remains negligibly small throughout space.

The quantum corrections to the metric outside the star are larger near the surface.
In order to study geodesics for which the quantum corrections are expected to be the largest,
we impose the boundary conditions
\begin{align}
	&
	r_{\rm c}(0)=\rs
	\ ,
	\\
&
r_{\rm q}\left(\lambda_0\to\infty\right) = 0
	\ .
	\label{rbinf}
\end{align}
This somewhat unconventional choice of specifying the boundary conditions at two different points is motivated by the fact that one cannot set
$r_{\rm q}(\lambda=0) = 0$, as the quantum corrections diverge at the surface of the star.
Instead one can impose any boundary condition on $r_{\rm q}( \lambda_0)$ for any $\lambda_0>0$, as this boundary condition does not impact the cumulative quantum corrections along a particular
segment of the geodesic. For this one has to evaluate the difference
$r_{\rm q}(\lambda_2) - r_{\rm q}(\lambda_1)$, for specified values $\lambda_1$ and $\lambda_2$, and any such difference is independent of the specific choice of $\lambda_0$.
\par
For $L=0$ one finds the leading classical solutions for an outgoing radial geodesic ($\lambda\geq0$)
\begin{align}
	r_{0,0}(\lambda) 
	&= \eta \,\lambda + \rs, \label{GeodesicExt_r00}
	\\
	r_{0,1}(\lambda) 
	&= \frac{\epsilon \,\rs}{2\,\eta^2}\,
	\ln\left(1 + \frac{\eta\,\lambda}{\rs}\right)
	\ ,
\end{align}
and the leading quantum corrections
\begin{align}
	r_{1,0}(\lambda) &= 0\\
	r_{1,1}(\lambda) 
	&= \frac{\talp E^2 \rs}{2\,\eta^2} 
	\left[ 2 \ln\left(\frac{2\rs + \eta\lambda}{\rs+\eta\lambda} \right)
	-2
	+ \frac{\eta \lambda}{\rs} \ln \left(1 + \frac{2\rs}{\eta\lambda} \right)
	\right]
	\nonumber\\
	&\quad 
	+ \frac{\tbet \rs}{4}
	\ln \left[\frac{\eta \lambda (2\rs + \eta \lambda)}{(\rs + \eta \lambda)^2} \right]. \label{eq:GeoExt}
\end{align}
Notice that $r_{11}(\lambda)$ contains a secular term proportional to $\tbet$ for $\lambda\rightarrow0$,
which was expected, and occurs within the interval of Eq.~\eqref{eq:ExclInt}.

However, the term proportional to $E^2$ never grows large even for $\lambda\sim 0$, 
and the violation to the equivalence principle therefore remains of order $(\lp/\rs)^2$ everywhere
in $r>\rs$.
\subsection{Interior region}
In the interior region we can write
\begin{align}
	\left( \frac{dr}{d\lambda}\right)^2 
	+ \frac{L^2}{r^2} 
	- \left[ \left(\frac{L^2}{r^2} + \epsilon \right) \left(\frac{r}{\rs}\right)^2 
	+ \frac{3E^2 (\rs^2-r^2)}{2\rs^2} \right]	\frac{2\gn M}{\rs} &\nonumber\\
	- \frac{3E^2 (11\rs^4 - 14\rs^2r^2 + 3r^4)}{16\rs^4} \left(\frac{2\gn M}{\rs}\right)^2
	+ V_q(r) 
	&= \eta^2,
\end{align}
where we again set $\eta = \sqrt{E^2 - \epsilon}$ and the effective quantum potential now
reads
\begin{equation}
	V_{\rm q}(r)
	=
	E^2 \, \alpha_{\rm i}(r) + \eta^2\, \beta_{\rm i}(r) - \frac{L^2}{r^2}\, \beta_{\rm i}(r)
	\ .
\end{equation}
Like in the exterior, we impose initial conditions suitable for studying radial geodesics near the
surface, that is
\begin{align}
	r_{\rm c}(0) &=\rs
	\ ,
	\\\
	r_{\rm q}\left(- \frac{\rs}{\eta}\right) &= A
	\ ,
\end{align}
where we will fix the value of $A$ at a later stage. 
For $L=0$ one finds the leading classical solution for an outgoing radial geodesic ($\lambda\leq0$)
is given by
\begin{align}
	r_{0,0}(\lambda) 
	&= \eta \, \lambda + \rs,\\
	r_{0,1}(\lambda) 
	&= \frac{\epsilon\, \lambda}{2\,\eta}
	- (3\, E^2 - 2\, \epsilon)\, (3\, \rs + \eta\,\lambda)\, \frac{\lambda^2}{12\, \rs^2}
	\ ,
\end{align}
and the leading quantum corrections read
\begin{align}
	r_{1,0}(\lambda) &= 0
	\\
	r_{1,1}(\lambda;x) 
	&= \frac{\talp E^2 \rs}{2\eta^2} 
	\left[ 2 \ln\left(2 + \frac{\eta\lambda}{\rs} \right)
	-2
	+ \frac{\eta \lambda}{\rs} \left\{ 
	\ln \left[- \frac{\eta\lambda}{\rs}\left(2 + \frac{\eta\lambda}{\rs}\right) \right] 
	-2 \right\}
	\right]\nonumber\\
	&\qquad
	+ \frac{\tbet \rs}{4} \,
	\left\{2 + 2\, \frac{\eta\lambda}{\rs} 
	- \ln \left[-\left(1+ \frac{2\rs}{\eta\lambda}\right) \right]
	\right\} + A\ . \label{eq:GeoInt}
\end{align}
Notice that $r_{11}(\lambda)$ also contains a secular term proportional to $\tbet$ for $\lambda\to 0$
which, like for the exterior expression~\eqref{eq:GeoExt}, occurs within the interval given in Eq.~\eqref{eq:ExclInt}.
\subsection{Crossing the surface}
By means of the previous results, we can analyze the discontinuities (of quantum origin) that the radial geodesics
would encounter across $r=\rs$.
Since we assumed initial conditions such that the classical radial geodesics $r_{\rm c}$ can be joined continuously
across $r=\rs$, we just need to calculate the difference between the non-vanishing quantum exterior correction in
Eq.~\eqref{eq:GeoExt} and the interior analogue in Eq.~\eqref{eq:GeoInt} at $r=\rs$, which yields
\begin{equation}
	\lim_{\lambda\rightarrow 0}\left[ 
	r^{\rm ext}_{1,1}(\lambda) 
	- r^{\rm int}_{1,1}(\lambda) \right]
	=
	\frac{\tbet\, \rs}{2} \,\left[
	\ln(2) - 1 \right] - A.
\end{equation}
We then notice that the interior and exterior geodesics can be continuously connected by fixing $A$ such that the
boundary condition for the interior solution is given by
\begin{equation}
	r^{\rm int}_{\rm q}\left(-\frac{\rs}{\eta}\right) 
	= 
	\frac{\tbet\, \rs}{2} \,\left[ \ln(2)  - 1 \right]
	\ ,
\end{equation}
provided for the exterior solution one employs the condition
\begin{equation}
	r^{\rm ext}_{\rm q}\left(\infty\right) 
	= 0
	\ ,
\end{equation}
which was used to determine Eq.~\eqref{eq:GeoExt}.
\par
One could go further and check the smoothness of the solution, and find that there is a discontinuity in the first derivative
that cannot be removed.
However, this is not a physical effect, as it occurs in the interval~\eqref{eq:ExclInt}, and is thus expected to be regularized
once the interior Schwarzschild solution is smoothened like we wrote in the Introduction.
\section{Scalar Fields}
\label{S:4}
The equation of motion for a free scalar field $\Phi$ with mass $\mu$ is given by
\begin{equation}
\label{eq:WaveEq}
	\Box\, \Phi = \mu^2\, \Phi
	\ .
\end{equation}
Since our metric~\eqref{eq:metricgen} has spherical symmetry, we can separate the angular variables
from the other coordinates and write $\Phi(t,r,\theta,\phi)=\Phi(t,r) \,S(\theta,\phi)$,
where $S$ can be decomposed in the usual spherical harmonics satisfying
\begin{equation}
\left(\partial_\theta^2 + \frac{\cos\theta}{\sin\theta} \,\partial_\theta + \frac{1}{(\sin\theta)^2}\,\partial_\phi^2 \right) Y(\theta,\phi)
=
-\,l(l+1)\, Y(\theta,\phi)
\ .
\end{equation}
\par
It is then convenient to consider one mode at a time and further separate time from the radial coordinate,
to wit $\Phi(t,r)=\Psi(t)\,\Phi(r)$, where $\Psi\sim e^{i\,\omega\,t}$ and satisfies
\begin{equation}
	\ddot{\Psi}(t) = - \omega^2\, \Psi(t)
	\ .
\end{equation}
Furthermore, using the metric~\eqref{eq:metricgen} with the tortoise-like coordinate $r_*$ yields
the radial equation
\begin{equation}
\label{eq:RadialEq}
	\left[\partial_{r_\ast}^2 + \eta^2 - \frac{l(l+1)}{r^2} \right] u(r)
	=
	\left(V_{\rm c} + V_{\rm q} \right) u(r)
	\ ,
\end{equation}
where $r_\ast$ is given as a function of $r$ in Eq.~\eqref{eq:rast},
$\eta^2 = \omega^2 - \mu^2>0$ and
\begin{equation}
	u(r)= r_\ast(r)\, \Phi(r)
	\ .
\end{equation}
Notice that we have explicitly separated the effective potential into a classical part,
\begin{equation}
\label{eq:Vc}
	V_{\rm c}(r) 
	= \left[f(r) - \alpha(r) - 1\right]
	\left[\mu^2 + \frac{l\,(l+1)}{r^2}\right]
\end{equation}
and a quantum contribution	
\begin{equation}
\label{eq:Vq}
V_{\rm q}(r) 
=  \alpha(r) \left[\mu^2 + \frac{l\,(l+1)}{r^2}\right]
\ .
\end{equation}
\par
Like for the geodesics, we can expand the radial function in the same perturbative parameters of
the quantum corrections to the metric and write
\begin{align}
u(r)
&=
u_{\rm c}(r)
+
u_{\rm q}(r)
\nonumber
\\
&=
\sum_{n,m=0}^{\infty} u_{n,m}(r) 
\left(\frac{\lp}{\rs}\right)^{2n}
\left(\frac{2\gn M}{\rs}\right)^{m}
\ ,
\end{align}
where $u_{\rm c}$ contains all the terms with $n=0$.
\par
We are particularly interested in how quantum corrections to the metric affect the $s$-waves with $l=0$
originating near the surface of the star.
On using the fact that $V_{\rm c}$ and $V_{\rm q}$ are of order (at least) $\gn$,
we immediately obtain
\begin{align}
\label{eq:S00}
	u_{0,0}(r) &= A \cos \left[\eta \,(r_\ast - \rsa) \right]
	\ ,
	\\
\label{eq:S10}
	u_{1,0}(r) &= 0
	\ ,
\end{align}
where $A$ and $\rsa$ are integration constants which we will suitably set in the following subsections.
The effect of the potentials~\eqref{eq:Vc} and~\eqref{eq:Vq} can then be determined perturbatively
by treating them as sources acting on the unperturbed solutions defined by Eqs.~\eqref{eq:S00} and~\eqref{eq:S10}.
\subsection{Exterior Region}
In the exterior region, the tortoise coordinate is given by
\begin{equation}
\label{eq:tortoiseOut}
	r_\ast(r) 
	= r + 2\,\gn M\ln\left( \frac{r}{2\,\gn M} - 1\right)
	+ \frac{1}{2} \int_{r}^\infty \left[\alpha_{\rm e}(r') - \beta_{\rm e}(r')\right]
	dr'
	+C
	\ ,
\end{equation}
where we set the integration constant $C$ so that
\begin{equation}
\rsa = \rs + 2\,\gn\, M\ln\left( \frac{\rs}{2\,\gn M} - 1\right)
\ .
\end{equation}
In order to determine the radial function in such a way that all corrections
to the unperturbed solutions~\eqref{eq:S00}
and~\eqref{eq:S10} vanish at some $r=(1+\delta)\,\rs>\rs$, we impose the boundary condition 
\begin{equation}
u[(1+\delta)\,\rs]
=
A
\ ,
\end{equation}
where $A$ is the same constant as in Eq.~\eqref{eq:S00} and $\delta$ is the same parameter
that defines the excluded layer in Eq.~\eqref{eq:ExclInt}.
\par
We want to see how these modes behave for values of $r> (1+\delta)\,\rs$.
The radial equation~\eqref{eq:RadialEq} can then be rewritten as the integral equation
\begin{equation}
	u(r_*) = A\, \cos [\eta\, (r_\ast - \rsa)]
	+ 
	\int_{(1 + \delta)\,\rsa}^{\infty}  G(r_\ast,r_\ast') 
	\left[ V_{\rm c}(r'_*) + V_{\rm q}(r'_*) \right]
	u(r'_*) \, dr_\ast'
	\ ,
\end{equation}
where the Green's function is given by
\begin{equation}
	G(r_\ast,r_\ast') = \begin{cases}
	\frac{1}{\eta} \sin\left[\eta (r_\ast - r_\ast') \right]   \quad
	& {\rm if} \quad r_\ast'\leq r_\ast, \\
	0 
	& {\rm if} \quad r'_\ast>r_\ast.	
	\end{cases}\\
\end{equation}
In order to solve the integral equation, one needs to invert Eq.~\eqref{eq:tortoiseOut}, which can be done
perturbatively using
\begin{equation}\label{eq:rastWaveExt}
	r_\ast - \rsa 
	= r - \rs  
	+ \mathcal {O}\left(2\,\gn M/\rs\right)
	\ .
\end{equation}
This is valid if the secularity is avoided, which is the case for $r>(1+\delta)\,\rs$.
The leading classical solution is then found to be~\footnote{Note we make use of Eq.~\eqref{eq:rastWaveExt}
also in order to express the result in the coordinate $r$.}
\begin{align}
	u_{0,1}(r) 
	&= \frac{ \mu^2 \rs A}{2\eta} \Big\{ 
	\ln(\rs/r) \sin\left[\eta (r_\ast - \rsa) \right]
	+ \left[ {\rm Si}(2\eta r) - {\rm Si}(2\eta \rs) \right]\,  \cos\left[\eta (r + \rs) \right]
	\nonumber\\
	&\qquad \qquad \quad
	- \left[ {\rm Ci}(2\eta r) - {\rm Ci}(2\eta \rs) \right]\, \sin\left[\eta (r + \rs) \right] \Big\},
\end{align}
and the leading quantum correction
\begin{align}
	u_{1,1}(r) 
	&= \frac{\talp \mu^2 A}{4\eta^2} \left(
	\left\{ \gamma_{\rm E}
	+ \ln\left[\frac{4 \eta \rs (r - \rs)}{r+\rs} \right] 
	- {\rm Ci}(2\eta |r-\rs|) \right\} \cos[\eta (r_\ast - \rsa)]
	\right. \nonumber \\
	& \qquad \qquad \quad
	+ \left[ 
	4\eta\rs 
	\ln\left(\frac{2 r}{r + \rs} \right) 
	+ 2\eta (r - \rs) 
	\ln\left(\frac{r - \rs}{r+\rs} \right) 
	\right. \nonumber\\
	& \qquad \qquad \qquad \qquad \qquad \qquad \qquad \qquad \qquad
	- {\rm Si}(2\eta |r-\rs|) \Big]
	\sin[\eta (r_\ast - \rsa)] 
	\nonumber \\
	& \qquad \qquad \quad
	- 4\eta\rs \left[{\rm Si}(2\eta r)
	- {\rm Si}(2\eta \rs) \right]\,  \cos[\eta (r + \rs)]
	\nonumber \\
	& \qquad \qquad \quad
	+ 4\eta\rs \left[ {\rm Ci}(2\eta r) 
	- {\rm Ci}(2\eta \rs) \right] \, \sin[\eta (r + \rs)]
	\nonumber \\
	& \qquad \qquad \quad
	+ \left\{ {\rm Ci}[2\eta (r+\rs)] 
	- {\rm Ci}(4\eta \rs) \right\} \, \cos[\eta (r + 3\rs)]
	\nonumber \\
	& \qquad \qquad \quad
	+ \left\{ {\rm Si}[2\eta (r+\rs)] 
	- {\rm Si}(4\eta \rs) \right\} \, \sin[\eta (r + 3\rs)] \Big),
\end{align}
where ${\rm Ci}$ and ${\rm Si}$ are cosine and sine integrals.
Notice that the results are independent of $\delta$, since corrections due to $\delta$ are subleading
by its definition in Eq.~\eqref{eq:ExclInt}.
Furthermore, at this order in perturbation theory, the divergences of the metric corrections~\eqref{eq:QcorOut}
can be absorbed in the phase
\begin{align}
	r_\ast - \rsa 
	&= r - \rs + 2\gn M \ln(r/\rs) + \frac{1}{2} \int_{r}^\infty \left[\alpha_e(r') - \beta_e(r')\right] dr' 
	\nonumber
	\\
	&= (r - \rs) \left( 1 + \frac{2\gn M}{\rs} \right)
	+ \left( \left\{ \talp\, [\ln(2) - 1] 
	+ \frac{\tbet}{4} \ln\left[ \frac{2(r-\rs)}{\rs}\right] \right\} \rs
	\right. \nonumber\\
	& \qquad \qquad \qquad \left.
	- \frac{1}{2} \left\{ \talp \left[ 1 + \ln\left(\frac{r-\rs}{2\rs} \right) \right]
	+ \frac{3}{4} \tbet \right\} (r - \rs)\right) \frac{2\gn M}{\rs} \frac{\lp^2}{\rs^2}
	\nonumber\\
	&\qquad
	+ \mathcal{O}\left(r-\rs\right)^2
	+ \mathcal{O}\left( \frac{2\gn M}{\rs} \right)^2
	+ \mathcal{O}\left( \frac{\lp}{\rs} \right)^4
	\ .
\end{align}
\subsection{Interior Region}
In the interior region the tortoise coordinate is given by
\begin{equation}\label{eq:tortoiseIn}
	r_\ast(r) 
	= r 
	+ \frac{r}{4} \left(3 - \frac{r^2}{\rs^2} \right) \frac{2\gn M}{\rs}
	- \frac{1}{2} \int_{0}^{r} \left[\alpha_i(r') - \beta_i(r')\right] dr'
	+
	D
	\ ,
\end{equation}
and $D$ is chosen so that
\begin{equation}
\rsa 
= \rs
+ {\gn M}
\ .
\end{equation}
We again impose a boundary condition, in order to fix the wave mode this time at $r=(1-\delta)\,\rs$,
to wit
\begin{equation}
u[(1-\delta)\,\rs]
=
A
\ .
\end{equation}
Like in the exterior, Eq.~\eqref{eq:RadialEq} yields the integral equation
\begin{equation}
	u(r) = A\, \cos [\eta \,(\rsa - r_\ast)] + 
	\int_{0}^{(1 - \delta)\,\rsa}  G(r_\ast,r_\ast') 
	\left[ V_c(r') + V_q(r') \right] u(r') \,dr_\ast'
	\ ,
\end{equation}
with the Green's function here given by
\begin{equation}
	G(r_\ast,r_\ast') = \begin{cases}
	0
	& {\rm if} \quad r_\ast' < r_\ast, \\
	\frac{1}{\eta}  \sin\left[\eta (r_\ast' - r_\ast) \right] \quad
	& {\rm if} \quad r'_\ast \geq r_\ast.	
	\end{cases}\\
\end{equation}
Eq.~\eqref{eq:tortoiseIn} can again be inverted perturbatively using
\begin{equation}
\label{eq:rastWaveInt}
	\rsa - r_\ast
	= \rs - r 
	+ \mathcal {O}\left(\frac{2\gn M}{\rs}\right)
	\ ,
\end{equation}
which is valid inside the ball $0\le r < (1-\delta)\,\rs$.
The leading classical solution is then found to be~\footnote{We make use of
Eq.~\eqref{eq:rastWaveInt} to revert to the coordinate $r$.}
\begin{align}
	u_{0,1}(r) 
	&= \frac{m^2 A}{24\eta^3 \rs^2} \Big\{ 
	3 \eta\, (r^2 - \rs^2) \cos[\eta(r_\ast - \rsa)]
	\nonumber \\
	& \qquad \qquad
	+ \Big[ 2\eta^2 ( r - \rs ) 
	( r^2 + r \rs - 8\rs^2 )
	- 3 ( r + \rs )\Big] \sin[\eta(r_\ast - \rsa)] \Big\}
	\ ,
\end{align}
and the leading quantum correction
\begin{align}
	u_{1,1}(r) 
	&= \frac{\talp m^2 A}{4\eta^2} \left(
	- \left[ \gamma_{\rm E}
	+ \ln\left(\frac{\eta |r^2 - \rs^2|}{\rs} \right)
	- {\rm Ci}(2\eta | r - \rs|) \right] \cos[\eta ( r_\ast - \rsa )]
	\right. \nonumber \\
	& \qquad \qquad \qquad
	+ \left\{ 
	4 \eta \rs \ln\left( \frac{2 \rs}{ r + \rs} \right) 
	+ 2\eta (r - \rs) \left[ 2
	- \ln\left(\frac{| r^2 -\rs^2|}{\rs^2} \right) \right]
	\right. \nonumber\\
	& \qquad \qquad \qquad \qquad \qquad \qquad \qquad \qquad \qquad \qquad 
	- {\rm Si}(2\eta | r - \rs|) \Big\} \sin[\eta (r_\ast - \rsa)] 
	\nonumber \\
	& \qquad \qquad \qquad
	+ \left\{ {\rm Ci}[2\eta (r + \rs)]   
	- {\rm Ci}(4\eta \rs) \right\} \cos[\eta (r + 3\rs)]
	\nonumber \\
	& \qquad \qquad \qquad
	+ \left\{ {\rm Si}[2\eta (r + \rs)]
	- {\rm Si}(4\eta \rs) \right\} \sin[\eta (r + 3\rs)] \Big)
	\ .
\end{align}
The results in the interior are again independent of $\delta$, to leading order, 
and the divergences of the metric corrections~\eqref{eq:QcorIn} can also be absorbed in the phase
\begin{align}
	r_\ast - \rsa 
	&= r - \rs
	+ \left[\frac{r}{4} \left(3 - \frac{r^2}{\rs^2} \right) - \frac{\rs}{2} \right] \frac{2\gn M}{\rs}
	- \frac{1}{2} \int_{0}^{r} \alpha_i(r') - \beta_i(r') dr' \nonumber\\
	&= r - \rs 
	- \left( \rs \left\{ \talp [1-\ln(2)] 
	+ \frac{\tbet}{4} \left[ 2 + \ln\left( \frac{|r-\rs|}{2\rs}\right) \right] \right\}
	\right. \nonumber\\
	& \qquad \qquad \left.
	+ \left\{\frac{\talp}{2}\left[1 - \ln\left(\frac{2|r-\rs|}{\rs} \right) 
	+ \frac{3\,\tbet}{8} \right] \right\} (r-\rs)\right) \frac{2\gn M}{\rs} \frac{\lp^2}{\rs^2}
	\nonumber\\
	&\qquad
	+ \mathcal{O}\left(r - \rs\right)^2
	+ \mathcal{O}\left( \frac{2\gn M}{\rs} \right)^2
	+ \mathcal{O}\left( \frac{\lp}{\rs} \right)^4
	\ .
\end{align}
\section{Discussion}
\label{S:5}
In this work we calculated the leading quantum corrections to the geodesics and the scalar waves
in a spacetime containing a constant and uniform density star.
We have shown as a proof of principle that such calculations can be done in quantum gravity.
Furthermore, we have found that the divergences at the surface of the star found
in Ref.~\cite{Calmet:2019eof}, do not cause serious issues for such calculations.
In fact, these divergences can be kept well under control, if a Planck length layer
around the surface of the star is excluded from the analysis.
It is then possible to connect the interior and exterior solutions in a continuous, but not differentiable
way, between the boundaries of such a layer.
\par
In the case of geodesics the quantum corrections only affect the velocity with respect
to the proper time for a particle following the geodesic.
For scalar waves on the other hand the quantum corrections give rise to both wavelike
perturbations to the classical wave solution and to a phase shift of the classical solution.
The latter could in principle lead to a measurable blueshift when the star surface is approached.
However, this would require compact objects to have density profiles that are smoothened
out within a Planck length interval around the surface of the star, and thus derivatives
of the energy density that exceed the Planck scale.
For any realistic matter distribution one would expect that all derivatives of the energy
density remain below the Planck scale.
\par
We conclude that neither the perturbations nor the phase shift are expected to be measurable
for realistic density profiles with current or near future experiments.
However, the latter effect is in fact very interesting, as it shows that quantum gravity introduces
a redshift due to the gradient of the density profile, while the redshift in general relativity results
only from the presence of mass.
\section*{Acknowledgments}
The work of X.C.~is supported in part  by the Science and Technology Facilities Council
(grant number ST/P000819/1).
R.C.~is partially supported by the INFN grant FLAG and his work has also been carried
out in the framework of activities
of the National Group of Mathematical Physics (GNFM, INdAM) and COST action {\em Cantata\/}. 
The work of F.K.~is supported by a doctoral studentship of the Science and
Technology Facilities Council.
F.K.~is grateful for the hospitality of the Universit\`a di Bologna, where most of this work
was carried out.
\end{document}